\documentstyle[twoside,fleqn,espcrc2,epsf]{article}

\newcommand{\half}{\mbox{\small $\frac{1}{2}$}}          % 1/2
\newcommand{\third}{\mbox{\small $\frac{1}{3}$}}         % 1/3
\def\lsim{\mathrel{\rlap{\lower4pt\hbox{\hskip1pt$\sim$}}
    \raise1pt\hbox{$<$}}}                % less than or approx. symbol
\def\gsim{\mathrel{\rlap{\lower4pt\hbox{\hskip1pt$\sim$}}
    \raise1pt\hbox{$>$}}}                % greater than or approx. symbol

\def\3{\ss}

\newcommand{\AmS}{{\protect\the\textfont2
  A\kern-.1667em\lower.5ex\hbox{M}\kern-.125emS}}

\title{
       \vspace{-3.65cm}                                     % for preprint
       {\normalsize DESY 97--177}    \\[-0.2cm]             % for preprint
       {\normalsize HUB--EP--97/57}  \\[-0.2cm]             % for preprint
       {\normalsize September 1997}  \\                     % for preprint
       \vspace{2.25cm}                                      % for preprint
       $O(a)$ Improvement of Nucleon Matrix Elements%
            \thanks{Talk given by R. Horsley at Lat97,      % for preprint
                    Edinburgh, U.K.}}                       % for preprint

\author{S.~Capitani%
           \address{Deutsches Elektronen-Synchrotron DESY,
                    D-22603 Hamburg, Germany},
        M.~G\"ockeler%
           \address{Institut f\"ur Theoretische Physik, Universit\"at
                    Regensburg, D-93040 Regensburg, Germany},
        R.~Horsley%
           \address{Institut f\"ur Physik, Humboldt-Universit\"at zu Berlin,
                    D-10115 Berlin, Germany},
        H.~Oelrich%
           \address{DESY-IfH Zeuthen, D-15735 Zeuthen, Germany},
        H.~Perlt%
           \address{Institut f\"ur Theoretische Physik, Universit\"at
                    Leipzig, D-04109 Leipzig, Germany},
        D. Pleiter$^{\rm d,}$
           \hspace{-0.2cm}
           \address{Institut f\"ur Theoretische Physik,
                    Freie Universit\"at Berlin, D-14195 Berlin, Germany},
        P.~E.~L. Rakow$^{\rm d}$,
        G.~Schierholz$^{\rm a,}$ \hspace{-0.2cm} $^{\rm d}$,
        A.~Schiller$^{\rm e}$
        and
        P.~Stephenson$^{\rm d}$}
       
\begin{document}

\begin{abstract}
We report on preliminary results of a high statistics quenched lattice QCD 
calculation of nucleon matrix elements within the Symanzik
improvement programme. Using the recently determined renormalisation
constants from the Alpha Collaboration we present a fully
non-pertubative calculation of the forward nucleon axial
matrix element with $O(a)$ lattice artifacts completely removed.
Runs are made at $\beta=6.0$ and $\beta=6.2$, in an attempt to check 
scaling and $O(a^2)$ effects. We shall also briefly describe
results for $\langle x \rangle$, the matrix element of a higher
derivative operator.
\end{abstract}

\maketitle

% reset footnote counter
\setcounter{footnote}{0}

%----------------------------------------------------------------------------

\section{INTRODUCTION}
\label{intro}

In this talk we shall describe results for nucleon matrix elements:
$\langle N|{\cal O}^R|N\rangle$ (at $\vec{p}=\vec{0}$) using $O(a)$
Symanzik improved fermions for
\begin{itemize}
   \item the vector current, $\langle N|{\cal V}^R_\mu|N\rangle$
         (as a warm-up exercise)
   \item the axial current,
         $\langle N,s|{\cal A}^R_\mu|N,s\rangle = \Delta q s_\mu$,
         where $\Delta q$ is the fraction of the nucleon spin carried by
         the quark $q$
   \item $\langle x\rangle^{(q)}$, the fraction of the nucleon
         momentum carried by quark $q$
\end{itemize}
We have worked in the quenched approximation
and generated $O(500)$ configurations at $\beta = 6.0$ on a
$16^3\times 32$ lattice and $O(150)$ configurations at
$\beta = 6.2$ on a $24^3\times 48$ lattice.
In both cases we have used three $\kappa$ values 
to enable us to extrapolate to the chiral limit. For the hadron
spectrum see \cite{stephenson97a}.
The method to determine the matrix elements is standard,
see eg \cite{goeckeler95a}; we only note that we are computing
just the quark line connected term.

\section{IMPROVED OPERATORS}
\label{operators}

Symanzik improvement is a systematic improvement of the action and
operators to $O(a^n)$ (here $O(a^2)$) by adding a basis of
irrelevant operators to completely remove $O(a^{n-1})$ effects.
Restricting improvement to on-shell matrix elements means
that the equations of motion ($EOM$) can be used to reduce the 
set of operators. For the action we only need one additional
operator -- the clover term, \cite{sheikholeslami85a}, with known 
coefficient $c_{sw}(g^2)$, \cite{luescher97a}.
We write for the improved axial, vector currents:
\begin{eqnarray}
   {\cal V}_\mu &=& V_\mu
                    - c_{1V}a\bar{q}\stackrel{\leftrightarrow}{D}_\mu q
                    + \ldots ,
                                         \nonumber  \\
   {\cal A}_\mu &=& A_\mu
                    - c_{1A}ai \bar{q}\sigma_{\mu\lambda}\gamma_5
                        \stackrel{\leftrightarrow}{D}_\lambda q
                    + \ldots ,
                                          \nonumber
\label{v+a_def}
\end{eqnarray}
while
$\langle N|{\cal T}_{44} - \third {\sum_i\cal T}_{ii} |N\rangle
 = -2m_N^2\langle x\rangle$, with
\begin{eqnarray}
   {\cal T}_{\mu\nu} = \begin{array}[t]{l}
                          \bar{q}\gamma_\mu
                             \stackrel{\leftrightarrow}{D}_\nu q
                             + c_1 ai \bar{q} \sigma_{\mu\lambda}
                          \stackrel{\leftrightarrow}{D}_{[\nu}
                          \stackrel{\leftrightarrow}{D}_{\lambda]} q  \\
                          -  c_2 a\bar{q}
                          \stackrel{\leftrightarrow}{D}_{\{\mu}
                          \stackrel{\leftrightarrow}{D}_{\nu\}} q
                          + \ldots ,
                        \end{array}
                                          \nonumber
\label{x_def}
\end{eqnarray}
and
$\stackrel{\leftrightarrow}{D}_\mu \equiv\half(\stackrel{\rightarrow}{D}_\mu -
\stackrel{\leftarrow}{D}_\mu)$. Derivative operators which do not
contribute to the forward matrix element have been dropped.
Renormalised operators are then given by
${\cal O}^R = Z_O {(1+b_Oam_q)\cal O}$. Note that these 
$O(a)$ operator sets with coefficients $b_O, \{c_I\}$ are over-complete
-- using the $EOM$ they can always be reduced by one.

\section{MATRIX ELEMENTS}

\begin{figure}[h]
   \vspace*{-1.00cm}
   \hspace*{-0.50cm}
   \epsfxsize=8.00cm \epsfbox{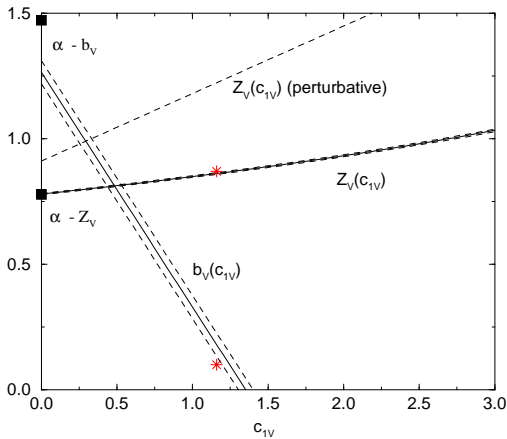}  % 10.0cm
   \vspace*{-1.25cm}
   \caption{\footnotesize
            Results for $Z_V$ and $b_V$ against $c_{1V}$
            for $\beta=6.0$. The error bars are given by bands of
            dotted lines. Also plotted is the one-loop perturbative
            result for $Z_V$ and the Alpha Collaboration results.}
   \vspace*{-0.75cm}
   \label{fig_Zv+bv}
\end{figure}

For the vector current we have from current conservation
$\langle N|{\cal V}_4^{(f)R}|N\rangle = \chi^{(f)}$, with $\chi^{(u)}=2$
and $\chi^{(d)}=1$. Using the $EOM$ we have a one parameter degree of
freedom, which we can take to be $c_{1V}$.
From linear fits (in $am_q \equiv \half (1/\kappa - 1/\kappa_c)$)
to $\langle N|V_4|N \rangle$,
$-a\langle N|\bar{q}\stackrel{\leftrightarrow}{D}_4 q|N\rangle$
we can find $Z_V$ from the constant terms and from the gradients $b_V$.
(Linear fits seem to be adequate within our quark mass range
$am_q \lsim 0.1$.) The result for $Z_V$, $b_V$ is shown in
Fig.~\ref{fig_Zv+bv} for $\beta=6.0$.
The Alpha Collaboration has non-perturbatively found
$Z_V(c_{1V}=0)$ and $b_V(c_{1V}=0)$, \cite{luescher97a}.
We see from the figure that there is very good agreement
between the different determinations for $Z_V(c_{1V}=0)$
and although for $b_V(c_{1V}=0)$ there is some discrepancy
we note that this is not unexpected as $b_V$ has $O(a)$
effects. Also shown are two points (stars) from a
method  described in \cite{rakow97a}. Again there is good agreement
(this time at a non-zero value of $c_{1V}$).

\begin{figure}[h]
   \vspace*{-1.00cm}
   \hspace*{-0.50cm}
   \epsfxsize=8.00cm \epsfbox{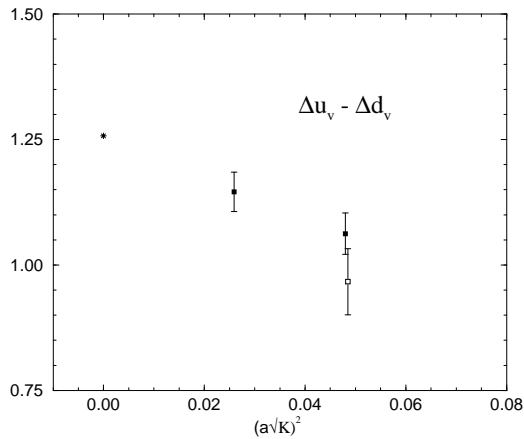}  % 10.0cm
   \vspace*{-1.25cm}
   \caption{\footnotesize
            A scaling plot of
            $\Delta u_v - \Delta d_v \equiv g_A$
            against $a^2$, using the string tension to set the scale.
            The Symanzik improved data at
            $\beta=6.0$, $6.2$ is shown with filled squares,
            the Wilson data, open square, is at $\beta=6.0$.
            The physical value of $g_A = 1.26$.}
   \vspace*{-0.75cm}
   \label{fig_ga_sig2}
\end{figure}

For the axial current, again the
Alpha Collaboration has non-perturbatively found $Z_A(c_{1A}=0)$,
\cite{luescher97a}. This is sufficient for us as we are only interested
in the results in the chiral limit. Performing these extrapolations
for $\beta=6.0$, $6.2$ gives the scaling plot, Fig.~\ref{fig_ga_sig2}.
To attempt a comparison with Wilson data, we have also plotted
the result at $\beta=6.0$, \cite{goeckeler95a}
using the non-perturbative $Z_A$ as found in \cite{rakow97a}
of $0.78$. This is close to the value ($0.75$) given
from linearly interpolating the non-perturbative Ward Identity
results in \cite{aoki97a} to $\beta = 6.0$, which would indicate 
that $O(a)$ effects between these two methods are small.
In Fig.~\ref{fig_ga_sig2} we expect to have to make a linear
extrapolation in $a^2$ for the improved case.

\begin{figure}[h]
   \vspace*{-1.00cm}
   \hspace*{-0.50cm}
   \epsfxsize=8.00cm \epsfbox{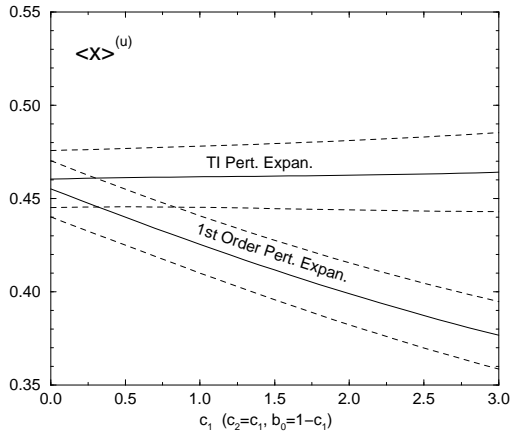}  % 10.0cm
   \vspace*{-1.25cm}
   \caption{\footnotesize
            A comparison between first order perturbation theory
            and $TI$ perturbation theory, using the boosted
            coupling $\alpha^{\overline{MS}}(\mu a)$ given in
            table $I$ of \cite{lepage93a}. Also used in the perturbative
            coefficient is $\tilde{c}_{sw}=c_{sw}(g^2)u_0^3$,
            ($u_0^4=\langle \third \mbox{tr} U_{plaq}\rangle$).}
   \vspace*{-0.75cm}
   \label{fig_x1b_b6p00c1p769+TI}
\end{figure}

\begin{figure}[h]
   \vspace*{-0.25cm}
   \hspace*{-0.50cm}
   \epsfxsize=8.00cm \epsfbox{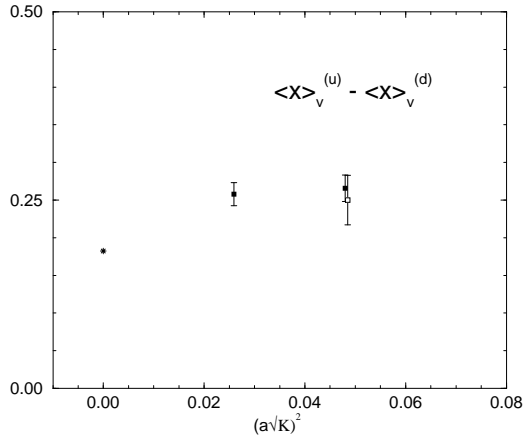}  % 10.0cm
   \vspace*{-1.25cm}
   \caption{\footnotesize
            Scaling plot of non-singlet
            $\langle x\rangle_v^{(u)} - \langle x\rangle_v^{(d)}$
            against $a^2$, using $c_1=0$.
            Same notation as in Fig.~\ref{fig_ga_sig2}.}
   \label{fig_x1b_u-d_sig2}   
   \vspace*{-0.75cm}
\end{figure}

For $\langle x \rangle$ there is no non-perturbative determination 
of the renormalisation constants yet available. First order perturbation
theory gives, however, numbers rather close to $1$, \cite{capitani97a},
and indeed using tadpole improved ($TI$) perturbation theory,
we see that with one derivative operators the (large) tadpole terms cancel.
This might indicate that perturbation theory is reasonably correct.
From the definition of the improved operator we see
that we have parameters $b_0$, $c_1$, $c_2$.
To remove $O(a)$ effects completely we must only have a residual one parameter
degree of freedom (from the $EOM$), so that, eg, $c_2 \equiv c_2(c_1)$
and $b_0 \equiv b_0(c_1)$. These functions are unknown at present.
At tree level $c_2=c_1$, $b_0=1-c_1$, which we shall
use here. In Fig.~\ref{fig_x1b_b6p00c1p769+TI}
we compare one loop perturbative results, and $TI$ results for
$\langle x\rangle^{(u)}$ in the chiral limit.
We know that upon using the true coefficients then the result
must be independent of $c_1$.
$TI$ appears to achieve this somewhat better than $1$-loop perturbation
expansion, so we shall use this in our scaling plot
shown in Fig.~\ref{fig_x1b_u-d_sig2}.
This operator has an anomalous dimension, so we
must scale the results to the same $\mu$; this has been
performed with the scaling formula
$\langle x \rangle |_\mu = (\alpha_{\overline{MS}}(\mu) /
\alpha_{\overline{MS}}(\mu_0))^{32/99}\langle x \rangle |_{\mu_0}$.
Scaling the $\beta=6.2$ result to $\beta=6.0$ where
$\mu \sim 1.95 (\mbox{GeV})^{-1}$ gives a $4\%$ increase.
We compare with the phenomenological value, \cite{martin95a}.
Our results seem rather constant (in $a^2$),
suggesting that even in the continuum limit the lattice 
value is too high.
We can only speculate on the discrepancy:
most likely this is due to a lattice problem --  quenching,
chiral extrapolation, $Z$ not accurately enough known
or perhaps there is a phenomenological
problem -- the fits are made from low $\mu^2$ so there might
be higher twist contributions, \cite{levin97a}.

\section*{ACKNOWLEDGEMENTS}
\label{acknowledgement}

The numerical calculations were performed on the
Quadrics QH2 at DESY-IfH. Financial support from the
DFG is gratefully acknowledged.

%----------------------------------------------------------------------------

\end{document}